\documentstyle[12pt]{article}
\begin{document}
\bibliographystyle{unsrt}
\rightline{IASSNS-HEP-00/11}
\rightline{February, 2000}
\bigskip
\centerline{THE EQUILIBRIUM DISTRIBUTION OF GAS MOLECULES} 
\centerline{ADSORBED ON AN ACTIVE SURFACE}
\medskip
\centerline{Stephen L. Adler}
\centerline{Institute for Advanced Study}
\centerline{Princeton, NJ 08540}
\bigskip
\centerline{Indrajit Mitra}
\centerline{Department of Physics, Jadwin Hall}
\centerline{Princeton, NJ 08544}
\bigskip
\leftline{Send correspondence to:~Indrajit Mitra~~~imitra@princeton.edu}

\newpage

\begin{abstract}
    We evaluate the exact equilibrium distribution of gas molecules adsorbed 
on an active surface with an infinite number of attachment sites. Our result 
is a Poisson distribution having mean $X = {\mu P P_s \over P_e}$, with $\mu$ 
the mean gas density, $ P_s$ the sticking probability, $P_e$ the evaporation 
probability in a time interval $\tau$, and $P$ Smoluchowski's exit probability 
in time interval $\tau$ for the surface in question. We then solve for the 
case of a finite number of attachment sites using the mean field 
approximation, recovering in this case the Langmuir isotherm.

\end{abstract}

\newpage

\newpage

\section{Introduction}

One of the models aimed at explaining the collapse of the wave function [1] 
predicts that the wave function of every system collapses to an eigenstate of 
the Hamiltonian in the energy basis in a time which depends on the energy 
spread of the wave packet. For a system including the measuring apparatus, 
relevant sources of energy fluctuations are thermal energy fluctuations and 
energy (mass) fluctuations coming from fluctuations in the number of surface 
adsorbed molecules. Our aim in this paper is to derive formulas for the 
equilibrium distribution of adsorbed molecules on an active surface S, from 
which the root mean square mass fluctuation can be calculated. Our results 
could also be relevant in other contexts - e.g. in surface catalysis.

Since our work is based largely on the classical colloidal statistics problem 
[2] solved by Smoluchowski, we will review his result first. Consider a gas 
chamber of volume $V$ which has $N$ gas molecules distributed randomly inside. 
Assuming uniform occupancy, the probability that a single molecule is found 
inside a small subvolume $v$ is ${v \over V}$ and of not being found inside is 
${V-v \over V}$. So the probability $U(n)$ of some $n$ particles being found 
inside $v$ is given by the binomial distribution
  
\begin{equation}
 U(n) = {N \choose n}({v \over V})^n({1 - {v \over V}})^{N-n}
\end{equation}
The mean number of particles $\mu$ found inside the small volume $v$ is just 
the mean of this binomial distribution ${Nv \over V}$. In terms of $\mu$ then, 
the distribution $U(n)$ becomes

\begin{equation}
U(n) = {N \choose n}({\mu \over N})^n({1- {\mu \over N}})^{N-n}
\end{equation} 
For most practical cases $N$ and $V$ are both very large, but the ratio of 
$N/V$ is finite so that the mean $\mu$ is finite. In this limit, the binomial 
distribution of Eq.(2) reduces to the Poissonian form

\begin{equation}
U(n) = {e^{-\mu}\mu^n \over n!}
\end{equation}

The interpretation of this equation is the following: If we focus on a small 
subvolume $v$ inside a much larger volume $V$, then the frequency with which 
different numbers of particles will be observed in the smaller volume will 
follow a Poisson distribution. It should be noted that in addition to the 
assumption of all positions in the volume having equal a priori probability of 
occupancy, we also assume that the motions of individual particles are 
mutually independent. In the surface adsorption generalization discussed in 
Sec. 2, this is the case for an infinite number of attachment sites, but would 
not be the case for a finite number of attachment sites.

Let us now define $P$ to be the probability that a particle somewhere inside 
the small volume $v$ will have emerged from it during the time interval 
${\tau}$. The ``probability after-effect factor'' $P$ will depend on physical 
parameters such as the velocity distribution and mean free path of the 
particles, as well as the geometry of the surface boundary. In terms of $P$, 
the probability that starting with an initial situation of $n$ molecules 
inside $v$, $i$ of them escape in time ${\tau}$ is

\begin{equation}
A(n,i)={n \choose i}P^i (1-P)^{n-i}
\end{equation}

Let ${E_i}$ denote the probability of the volume $v$ 
capturing $i$ particles during time ${\tau}$. ${E_i}$ clearly is independent 
of the number of molecules already inside. But, under equilibrium conditions, 
the a priori probabilities for entrance and exit must be equal. For each $n$ 
there is a contribution to the exit probability; summing over all of them and 
equating to $E_i$ we get

\begin{equation}
E_i = \sum_{n=i}^{\infty} U(n)A(n,i)
\end{equation}
 Inserting the expressions for $U(n)$ and $A(n,i)$ from Eqs.(3) and (4) we get

\begin{equation}
E_i = \sum_{n=i}^{\infty} {e^{- \mu}\mu ^n \over n!} {n \choose i} P^i 
(1-P)^{n-i} = {e^{-\mu}(\mu P)^i \over i!} \sum_{n=i}^{\infty} 
{\mu^{n-i}(1-P)^{n-i} \over {(n-i)!}} = \sigma(i,\mu P)
\end{equation}
\\
where from here on we denote a Poisson distribution with mean $X$ by 
$\sigma(n,X)$ with
\begin{equation}
\sigma(n,X) = {{e^{-X} X^n} \over {n!}}
\end{equation}

\section{Adsorption of gas molecules}

To make our analysis intuitively clear, let us draw an imaginary surface $I$ 
just outside  the active surface area $S$. The following notations will be 
used:
(i) $E_i$ = Probability for $i$ molecules to enter the volume enclosed by 
$I$ in the time inverval $\tau$. Since this is the same as in the case where 
the surface $S$ inside is absent, this probability is just as in Eq.(6),

\begin{equation}
E_i =  \sigma (i,\mu P)
\end{equation}
\\
(ii) $U(n)$ = Probability to observe $n$ molecules sticking to $S$.\\
(iii) $P_s$ = Probability of a molecule to stick to $S$ after 
crossing $I$.\\
(iv) $P_e$ = Probability for a molecule that is stuck to $S$ to evaporate off 
in a time interval $\tau$.\\
(v) $B(n,i)$ = Probability that starting with an initial situation with 
$n$ particles stuck to $S$, $i$ of them evaporate in time $\tau$.\\

By Smoluchowski's reasoning leading to Eq.(4) above, we have 

\begin{equation}
B(n,i) = {n \choose i} (P_e)^i (1-P_e)^{n-i}
\end{equation}
 At equilibrium, the detailed balance condition holds. This is just the 
condition that the  probability that
$i$ particles stick in a time interval $\tau$ is equal to the probability that 
$i$ particles evaporate in the same time interval $\tau$.
The probability for $i$ molecules to stick to $S$ is $$ \sum_{j \geq i} E_j {j 
\choose i} P_s ^i (1-P_s)^{j-i}$$ Using Smoluchowski's  expression for $E_j$ 
from Eq.(8) this becomes
\begin{equation}
e^{-\mu P} { (\mu P P_s)^i \over i!} \sum_{j \geq i} {[\mu P (1-P_s)]^{j-i} 
\over (j-i)!} = \sigma(i,\mu PP_s)
\end{equation}   
\\

The other part of the detailed balance condition, the probability that out of 
$n$ molecules on $S$, $i$ of them evaporate in time interval $\tau$  is
\begin{equation}
 \sum_{n \geq i} U(n)B(n,i) = \sum_{n \geq i} U(n) {n \choose i} P_e^i 
(1-P_e)^{n-i}.
\end{equation}
\\
Equating these two probabilities, we have 
\begin{equation}
 \sigma(i,\mu PP_s) = \sum_{n \geq i} U(n) { n \choose i} P_e ^i (1-P_e)^{n-i}.
\end{equation}
\\

Our task now is to determine the equilibrium distribution $U(n)$ from this 
equation. We start with the ansatz that $U(n)$ is a Poisson distribution 
$\sigma(n,X)$ with a mean $X$ which is to be determined,

\begin{equation}
U(n) = \sigma(n,X) = {e^{-X} X^n \over {n!}}
\end{equation}
\\
Substituting Eq.(7) into Eq.(12) and using the sum evaluated in Eq.(6), we get 
the condition

\begin{equation}
\sigma(i,\mu P P_s) = \sigma(i,X P_e)
\end{equation}
\\
which is satisfied when

\begin{equation}
X = {\mu PP_s \over P_e}
\end{equation}
\\
Eqs.(13) and (15) are our result for the equilibrium distribution of adsorbed 
molecules. We note that, as intuitively expected, the mean number of adsorbed 
molecules increases with increasing gas density $\mu$ and increasing sticking 
probability $P_s$, but decreases with increasing evaporation probability $P_e$.

As a check on our reasoning, let us calculate the transition probability 
$W(n,m)$ for $m$ particles to be stuck to the surface at time $T+\tau$ when 
$n$ particles were stuck to the surface at time $T$, and then check that 
$W(n,m)$ and $U(n)$ have the requisite Markoff property.
The transition probability is given by 
\begin{equation}
W(n,m) = \sum_{x+y=m} W_{1}^{(n)}(x) W_2(y)
\end{equation}
\\
where $W_{1}^{(n)}(x)$ is the probability that $x$ particles remain at time 
$T+\tau$ when initially there were $n$ at time $T$,
\begin{equation}
W_{1}^{(n)}(x) = {n \choose x} (1-P_e)^{x}P_e^{n-x}
\end{equation}
\\
and $W_2(y)$ is the probability for $y$ additional particles to adhere to the 
surface in time ${\tau}$ as given by Eq.(10)
\begin{equation}
W_2(y) = \sigma(y,\mu P P_s)
\end{equation}

The Markoff property requires that 
\begin{equation}
U(m) = \sum_{n} U(n)W(n,m)
\end{equation}
\\
with $U(m)$ the equilibrium distribution of Eqs.(13) and (15). Evaluating the 
sum on the right-hand side of Eq.(19), we find as required that $$\sum_{x+y=m} 
(\sum_{n} U(n) W_{1}^{(n)}(x)) W_2(y) = \sum_{x+y=m} \sigma(x,X(1-P_e)) 
\sigma(y,\mu P P_s)$$ $$= \sigma(m,X(1-P_e)+\mu P P_s) = \sigma(m,X) = U(m)$$

\section{Finite number of attachment sites - mean field approach}

Let us now proceed to calculate the equilibrium distribution of the number of 
molecules attached to $S$ where $S$ has a finite (although very large) number 
of attachment sites $M$. Clearly, our discussion of the previous section 
breaks down since the sticking probability is no longer
a constant, but depends on the number $n$ of molecules already attached to 
$S$. In the following discussion, let us use $P_s$ to denote the probability 
for a molecule to stick to $S$ if no site is occupied, and let us denote the 
mean number of occupied sites by ${\overline m}$. Then the mean sticking 
probability is just

\begin{equation}
{\overline P_s} = P_s (1- {\overline m \over M})
\end{equation}
\\
and the corresponding distribution of stuck molecules is $\sigma(n,{\overline 
X})$, with

\begin{equation}
\overline X = {\mu P \overline P_s \over P_e}
\end{equation}
\\
Since the mean of this distribution is ${\overline m} = {\overline X}$, we get 
the mean field consistency condition

\begin{equation}
{\overline m} = (1- {\overline m \over M}){\mu PP_s \over P_e}
\end{equation}
\\
with solution
\begin{equation}
{\overline m} = {\mu PP_s/P_e \over (1+ {\mu PP_s \over P_e M })}
\end{equation}
\\
Thus, the mean fraction ${{\overline m} \over M}$ of total available sites 
occupied has the form of the Langmuir isotherm [3].

The mean field approximation is valid as long as the mean number of vacant 
sites $M - {\overline m}$ is much larger than the width $\sqrt{\overline m}$ 
of the distribution of adsorbed molecules,

\begin{equation}
M - {\overline m} = M (1- {{\overline m} \over M}) >> \sqrt{\overline m} = 
\sqrt{M} \sqrt{{\overline m} \over M}
\end{equation}
\\
Close to saturation, when ${{\overline m} \over M} \approx 1$, substituting 
Eq.(23) into Eq.(24) gives the condition

\begin{equation}
1+ {{\mu P P_s} \over {P_e M }} << \sqrt M
\end{equation}
\\
which when $X={\mu P P_s \over P_e} >> M$ simplifies to 

\begin{equation}
X << M^{3 \over 2} 
\end{equation}

\section{Acknowledgments}

 	One of us (S.L.A) was supported in part by the Department of Energy under 
Grant No.  DE-FG02-90ER40542. He also wishes to thank J. Lebowitz, S. Redner, 
and R. Ziff for helpful e-mail correspondence.

\section{References}

[1]  See, e.g., N. Gisin, Helv. Phys. Acta  62, 363 (1989); 
I. C. Percival, Proc. R. Soc. Lond. A447, 189 (1994); 
L. P. Hughston, Proc. R. Soc. Lond. A452, 953 (1996).\hfill\break
\medskip
[2] M. v. Smoluchowski, Physik. Zeits. 17, 557 (1916) and 17, 585 (1916).We 
follow the exposition given in the review of S. Chandrasekhar,
Rev. Mod. Phys. 15, 1 (1943).\hfill\break
\medskip
[3] R. H. Fowler, ``Statistical Mechanics'', Cambridge, 1936, pp. 
828-830; R. H. Fowler and E. A. Guggenheim, ``Statistical Thermodynamics'', 
Cambridge, 1939, pp. 426-428.\hfill\break

\end{document}